\documentclass{vgtc}                          

\ifpdf
  \pdfoutput=1\relax                   
  \usepackage{graphicx}                
  \DeclareGraphicsExtensions{.pdf,.png,.jpg,.jpeg} 
\else
  \usepackage{graphicx}                
  \DeclareGraphicsExtensions{.eps}     
\fi%

\graphicspath{{figures/}{pictures/}{images/}{./}} 

\usepackage[table]{xcolor}

\usepackage{microtype}                 
\PassOptionsToPackage{warn}{textcomp}  
\usepackage{textcomp}                  
\usepackage{mathptmx}                  
\usepackage{times}                     
\usepackage{cite}                      
\usepackage{tabu}                      
\usepackage{booktabs}                  

\usepackage{topcapt}
\usepackage[super]{nth}

\newcommand{\parahead}[1]
  {%
   \vspace{0.07in}%
   \noindent%
   \textbf{\textit{#1.}}%
  }
\newcommand{\etal}
  {et al.}
\newcommand{\etals}
  {et al.'s}  

\newcommand{\figref}[1]{\hyperref[#1]{Fig.~\ref*{#1}}}
\newcommand{\secref}[1]{\hyperref[#1]{Section~\ref*{#1}}}
\newcommand{\tabref}[1]{\hyperref[#1]{Table~\ref*{#1}}}

\newcommand{\midsepremove}{\aboverulesep = 0.0mm \belowrulesep = 0.0mm}
\definecolor{tableColor}{HTML}{daedf7}

\usepackage{wasysym}
\newcommand{\myCirc}[1]{{\large\color{#1}\CIRCLE}}
\definecolor{RomColor}{HTML}{62ACFF}
\newcommand{\Rom}{\myCirc{RomColor}}
\definecolor{ClasColor}{HTML}{00B318}
\newcommand{\Clas}{\myCirc{ClasColor}}
\definecolor{EmpColor}{HTML}{FFFF03}
\newcommand{\Emp}{\myCirc{EmpColor}}
\definecolor{InducColor}{HTML}{BA00B4}
\newcommand{\Induc}{\myCirc{InducColor}}
\definecolor{DeducColor}{HTML}{80FF00}
\newcommand{\Deduc}{\myCirc{DeducColor}}
\definecolor{RatioColor}{HTML}{FFA2E1}
\newcommand{\Ratio}{\myCirc{RatioColor}}
\definecolor{AnlgColor}{HTML}{FF0011}
\newcommand{\Anlg}{\myCirc{AnlgColor}}
\definecolor{MethodColor}{HTML}{A9FFD2}
\newcommand{\Method}{\myCirc{MethodColor}}
\definecolor{HistColor}{HTML}{1100FF}
\newcommand{\Hist}{\myCirc{HistColor}}
\definecolor{PhilColor}{HTML}{CC99FF}
\newcommand{\Phil}{\myCirc{PhilColor}}
\definecolor{MetaphColor}{HTML}{FFAE5C}
\newcommand{\Metaph}{\myCirc{MetaphColor}}
\definecolor{AgncyColor}{HTML}{9F5103}
\newcommand{\Agncy}{\myCirc{AgncyColor}}
\definecolor{FutColor}{HTML}{FF007F}
\newcommand{\Fut}{\myCirc{FutColor}}
\definecolor{ClsfColor}{HTML}{666600}
\newcommand{\Clsf}{\myCirc{ClsfColor}}
\definecolor{NumColor}{HTML}{000000}
\newcommand{\Num}{\myCirc{NumColor}}
\definecolor{WrColor}{HTML}{808080}
\newcommand{\Wr}{\myCirc{WrColor}}
\definecolor{BlColor}{HTML}{FFFFFF}
\newcommand{\Bl}{\myCirc{BlColor}}

\onlineid{0}

\vgtccategory{Research}

\vgtcinsertpkg

\title{Supporting Expert Close Analysis of Historical Scientific Writings:\\ A Case Study for Near-by Reading}

\author{
Andrew McNutt\thanks{The two first authors contributed equally to this work.} %
\and Agatha Kim$^*$
\and Sergio Elahi %
\and Kazutaka Takahashi %
}
\affiliation{
    \scriptsize University of Chicago
    \thanks{Email: \{mcnutt, agatha, sergioelahi, kazutaka\}@uchicago.edu}
}

\abstract{
Distant reading methodologies make use of computational processes to aid in the analysis of large text corpora which might not be pliable to traditional methods of scholarly analysis due to their volume. 
While these methods have been applied effectively to a variety of types of texts and contexts, they can leave unaddressed the needs of scholars in the humanities disciplines like history, who often engage in close reading of sources.
Complementing the close analysis of texts with some of the tools of distant reading, such as visualization, can resolve some of the issues. 
We focus on a particular category of this intersection---which we refer to as near-by reading---wherein an expert engages in a computer-mediated analysis of a text with which they are familiar. 
We provide an example of this approach by developing a visual analysis application for the near-by reading of \nth{19}-century scientific writings by J. W. von Goethe and A. P. de Candolle. 
We show that even the most formal and public texts, such as scientific treatises, can reveal unexpressed personal biases and philosophies that the authors themselves might not have recognized.
}

\CCScatlist{
  \CCScatTwelve{Applied computing}{Arts and humanities}{};
  \CCScatTwelve{Human-centered computing}{Visualization application domains}{}
}

\begin{document}


\firstsection{Introduction}

\maketitle

Historical analysis can be carried out using a variety of methods and techniques, depending on the types of the historical evidence at hand and the purpose of the investigating scholar.
As many historical sources are in written form, a predominant approach to their study is close reading, a process in which a scholar carefully studies a particular text to identify its message and intent.
In contrast to this approach is distant reading, a family of computer-mediated processes and techniques in which automated analysis methods, such as visualization and natural language processing, are applied to larger corpora \cite{moretti2005graphs}.
These methods are typically employed when the subject of concern is too large for traditional close reading.
Hope and Witmore\cite{hope2010hundredth} note that these computational prostheses can ``extend [text analysis] to where human readers simply can not go''.
Many works fruitfully employ these approaches,
however these techniques have been met with a mixed, or even contentious,
reception \cite{eve2019close, da2019computational}.

Augmenting close reading through visualization is a natural way to combine these streams of thought while avoiding some of the pitfalls endemic to computational studies\cite{da2019computational}.
Recent analyses of poetry \cite{mccurdy2015poemage}, rhetoric of Shakespeare's plays\cite{correll2011exploring}, and a variety of other topics \cite{campbell2018close, janicke2017visual} exhibit a combination of techniques that is not well captured by distant or close reading individually.
We argue that---by identifying distant and close reading as opposing poles in a spectrum of methods mediated by agency---they occupy their own unique space: \emph{near-by reading}.
In this mode of analysis, tools from distant reading are applied to a text with which the researcher is already familiar, in a manner that privileges both the specific content of the text and the automatically generated analyses and visualizations. We believe that this shifted perspective gives opportunity for novel analysis and discovery.

To investigate this notion, we conducted a collaborative design study with a historian---one of the authors---to see how visualization methods could aid a particular close reading task.
By comparing the scientific writings of \nth{19}-century authors J. W. von Goethe (1749-1832) and A. P. de Candolle (1778-1841), we highlighted their personal biases and philosophies  when conducting scientific research.
To make these texts tractable we constructed a manual multi-class categorization of each sentence in four representative texts.
We aimed to identify and analyze the distribution of categories in each scientific text, and investigate the interaction among key categories.
To serve these goals we created a web-based application which both enables explorations of the texts and introduces the project as a whole.
This application is available at \url{https://goetheanddecandolle.rcc.uchicago.edu/}.
This process revealed how these historical authors thought and expressed science and therein highlights the true differences between their scientific styles.
%

\newcommand{\halfLine}[1]{\raisebox{-0.65em}{#1}}
\midsepremove
\begin{table*}[t]
\centering
\rowcolors{2}{white}{tableColor}
\topcaption{
Each of the four texts relevant to our project was broken into sentences, which were tagged with one or more of the 17 categories that characterize the types of thinking relevant to our study. Here we show the categorization along with sentences that exemplify each category.
}
\small
\begin{tabular}{>{\raggedleft}p{.25in}>{\raggedleft}p{.95in}p{5.3in}}\toprule
  & Category & Example Sentence \\\midrule
\halfLine{1 \Rom} & \halfLine{Romantic} & There is a hidden relationship among various external parts of the plant which develop one after the other … The process by which one and the same organ appears in a variety of forms has been called the metamorphosis of plants.---\textbf{Goethe} \\
\halfLine{2 \Clas} & \halfLine{Classical} & It often happens that such plant which is thought to be an exception to a family or genera in which one has placed it, really belongs to a different family, when its organization is better known.---\textbf{DC1} \\
\halfLine{3 \Emp} & \halfLine{Empirical} & The [Ranunculus aquaticus] leaves produced underwater consist of threadlike ribs, while those developed above water are fully anastomosed and form a unified surface.---\textbf{Goethe} \\
\halfLine{4 \Induc} & \halfLine{Inductive} & If we still cannot state with certainty about such general relations between the secondary characters of fructification and those of nutrition, we nonetheless see them from numerous examples, to conclude that these relations truly exist.---\textbf{DC1} \\
\halfLine{5 \Deduc} & \halfLine{Deductive} & … proceeding from the opinion that the primitive nature is symmetrical, that irregularity is the product of diverse causes which alter this symmetry, we conceive that the monstrosities are due to certain variations of these causes.---\textbf{DC3} \\
\halfLine{6 \Ratio} & Rational or Speculative
(sentences not in 4 or 5) & This phenomenon can be due to either the multitude of islands which are dispersed in this sea, or to the fact that it has been traveled by navigators longer than any other, or maybe that it originated from some eruption of the ocean after the origin of the vegetation.---\textbf{DC2} \\
\halfLine{7 \Method} & \halfLine{Methodological} & … in [monstrosities] we see what organs are when they are not united together; we recognize what they are truly like when an accidental cause does not prevent them from developing.---\textbf{DC3} \\
\halfLine{8 \Hist} & \halfLine{Historical or Descriptive} & The ancient naturalists neglected the study and even the indication of homelands of plants. Linnaeus is the first who thought of indicating them in general works.---\textbf{DC2} \\
\halfLine{9 \Phil} & \halfLine{Philosophical} & … one must agree that the laws that are labeled as a priori can only be considered as more or less ingenious hypotheses, as long as they are not confirmed by observation.---\textbf{DC3} \\
\halfLine{10 \Anlg} & \halfLine{Analogical} & Thus the lateral branches stemming from the nodes of the plant can be considered as small plants attached to the parent in the same way that the parent is attached to the earth.---\textbf{Goethe} \\
\halfLine{11 \Metaph} & \halfLine{Metaphorical or Visual} & Species are large villages; genera are provinces; families are empires; classes are analogous to sections of the world, and the plants that remain isolated are represented by the islands distant from any continent.---\textbf{DC1} \\
\halfLine{12 \Agncy} & Metaphors attributing agency to nature & … nature has placed in the order of beings the empty spaces, here and there, just as she placed on the globe the inhabitable marshes and deserts.---\textbf{DC1} \\
13 \Clsf & Classificatory & In all bulbs, one distinguishes three parts: the radicles which emerge from underneath of the bulb, and which are true roots.---\textbf{DC1} \\
14 \Num & Numerical & … among 1485 vascular plants which grow in the British Isles, there are only 43 or 1/34 which have been found also in France.---\textbf{DC2} \\
\halfLine{15 \Fut} & Future researches \\or Utility & … a collection of properly arranged illustrations with the botanical terms for the different parts of the plant would be both pleasant and useful.---\textbf{Goethe} \\
\halfLine{16 \Wr} & Research goals \\or Directions & The aim of these observations is not to disrupt the classifications made by earlier observers and taxonomists; we only wish to clarify the variations in plant form.---\textbf{Goethe} \\
17 \Bl & Blank statements & Here is the result of this comparison.---\textbf{DC2} \\
\bottomrule
\end{tabular}
\label{tab:categories}
\end{table*}

\begin{figure*}[t] 
  \centering
  \includegraphics[width=\linewidth]{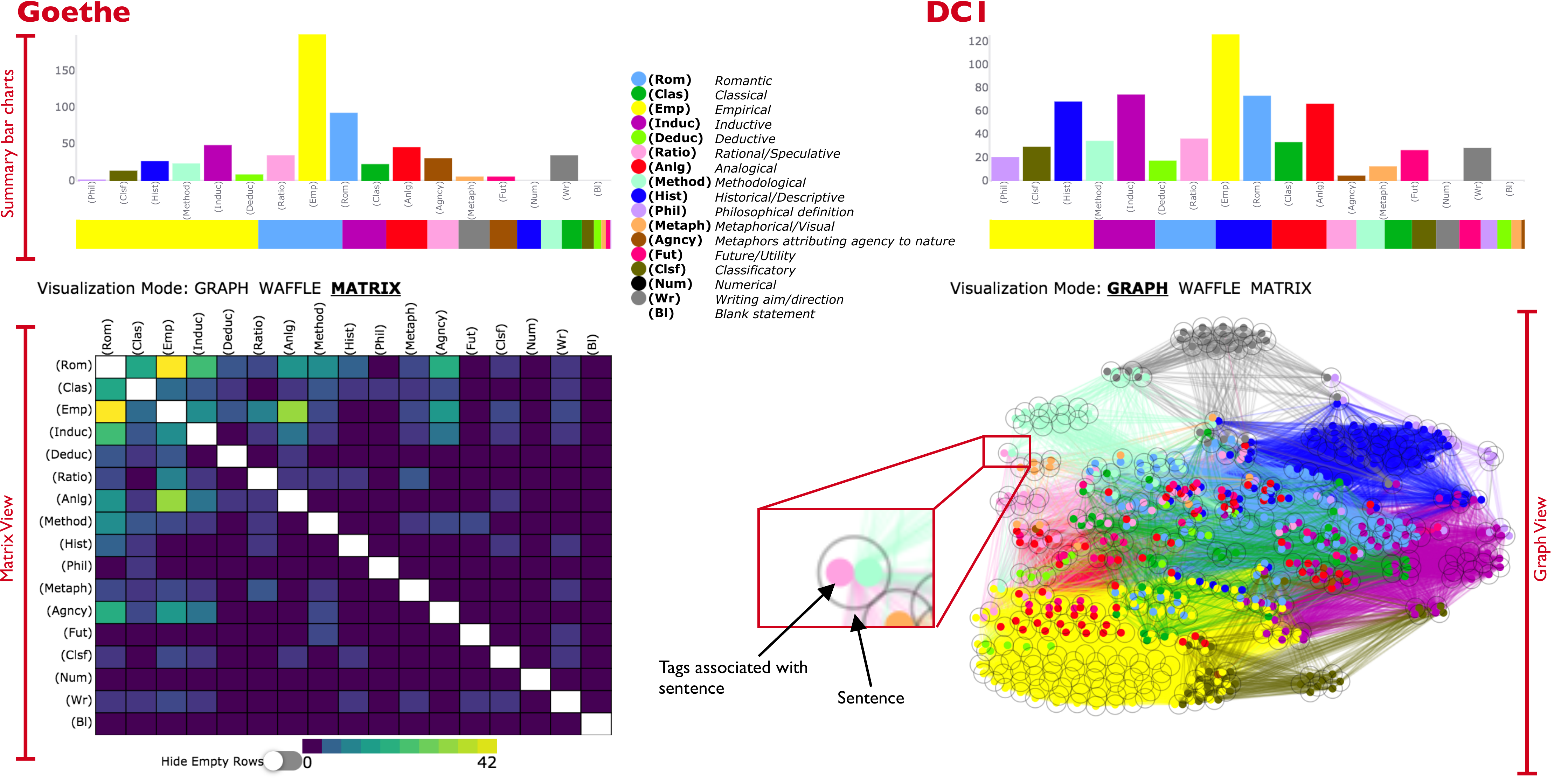}
  \caption{An annotated view of our application. The left column shows the matrix view of \textbf{Goethe}, while the right column displays \textbf{DC1} through the graph view (where the \emph{Blank}\Bl{} statements have been removed). Each sentence in the graph view is denoted by a gray ring, and each of its tags are shown as colored circles. These views enable comparison of modes of thinking exhibited by these authors and texts. }
  \vspace{-1em}
  \label{fig:application}
\end{figure*}

\section{Near-by Reading}\label{sec:nearby-reading}

Close and distant reading appear diametrically opposed.
Scholars engaging in close reading exercise full agency over their analysis, while in distant reading they only exercise agency in proxy, interacting with the underlying text only through the lens of their chosen automation; via data derived from the original text, such as counts, topics, entities, or a variety of other computationally constructible quantities.
Moretti argues this point explicitly, asserting that distant reading is the opposite of close reading \cite{moretti2005graphs}.

Yet the analysis process is rarely so cut-and-dry.
Some analyses lean more towards distant reading---only concerning themselves with data derived from the original text---while others weave together both derived data and inspection on the text itself.
J{\"a}nicke\cite{janicke2017visual} notes a history of combining close reading and distant reading techniques, from both top-down (privileging visualization) and bottom-up (privileging text) perspectives.
Campbell \etal~\cite{campbell2018close} employ a blended analysis that combines elements of distant and close reading, as do other works \cite{correll2011exploring, koch2014varifocalreader}.
Given the ambiguity of the boundary between these modes, we argue that they form a larger structural whole. In particular, we see these two approaches as poles in a spectrum, mediated by agency.
Each analysis falls somewhere in this spectrum depending on the roles carried out by the scholar and the machine.

Within this spectrum we identify distinct regions, based on intent, corpus volume, and exercised agency.
We are particularly interested in the region in which a scholar computationally analyzes texts that they have already investigated through traditional methods. We refer to this regime as \emph{Near-by reading}:
\begin{quote}
  \textbf{Near-by reading}: \emph{The computationally aided analysis of a text with which the analyst is already familiar.}
\end{quote}
This mode can be seen as close reading but at a distance, or a distant reading in which the guises of distance are made transparent.
A variety of prior works falls into this regime.
McCurdy \etal\cite{mccurdy2015poemage} ask several poets to use their tool to examine their own work---a near-by reading.
Eve analyzes the textual differences between the sections of Mitchell's \textit{Cloud Atlas}, a distant reading predicated on a deeply close reading---UK copyright law required Eve to literally retype the book in order to perform his computational analyses \cite{eve2019close}.
These approaches offer experts of a particular text a new vantage from which to make analyses.
We demonstrate the value of this regime through a case study in which we use visualization to aid in the analysis of historical scientific texts.
%

We believe that this lifted perspective, in which both poles of textual analysis are available, offers useful ground for analysis and reflection.
For instance, the insights gained from close reading and distant reading can create a dissonance which can provide valuable purchase for reflection and study.
Froelich\cite{froehlich2019Moby} points out counting words of Shakespeare's plays reveals that the word \emph{she} is relatively under-used in Macbeth compared to his other plays. This fact often surprises Shakespeare scholars, as that play is often perceived as being focused on Lady MacBeth.
Hope and Witmore use computers as prosthesis, furthering the abilities of the analysts by yielding the parts of the process where humans are lacking to computers (studious repetitive analysis) so as to aid those where we are strong (making abstract and nuanced connections).
They use these enhancements to analyze genre across Shakespeare's plays, which they in turn use as the foundation of subsequent analyses about the plays themselves (a similar sequence drives El-Assady \etals \cite{el2016visual} Knowledge Generation Loop).
Our near-by reading takes the opposite direction: previously analyzed text is externalized to gain a new view.

\section{Previous Work}

Applying visualization techniques to close reading has been investigated in previous studies, such as McCurdy \etals~\cite{mccurdy2015poemage} work on poetry, Campbell \etals~\cite{campbell2018close} work on analyzing TEI data, Gazzoni \etals~\cite{gazzoni2017mapping} work on mapping the Dante's Divine Comedy, and a variety of other contexts \cite{janicke2017visual}.
Edelstein \etal\cite{edelstein2018england} challenge the long-held view that the \nth{18}-century French philosopher Voltaire was an important propagator of the English ideas into France. This traditional view is based on  close readings of Voltaire's few writings that expressed admiration for English political and scientific thoughts. By visualizing Voltaire's actual correspondences with English figures, Edelstein \etal{} found that his English network was limited. Complementing this discovery with further close reading, they offer a more refined view that Voltaire was interested in the England of the past which had a strong tie to France, not in then-contemporary England---thus re-evaluating Voltaire's cosmopolitanism.
Like them, we embrace computational techniques in conjunction with close reading to make a useful contribution to the humanities research.
%

Our project is also related to Correll \etals~\cite{correll2011exploring} work on understanding tagged texts, although ours is less automated and on a smaller scale. Like them, we use a dimensionality reduction scheme in our graph view and provide details on demand controls; however, our project requires a different visual approach to accommodate our multi-class structure.
Our approach of surfacing set-like comparisons in our visualizations is preceded by a variety of prior work on comparing sets both in visualization \cite{lex2014upset, riche2010untangling} and in digital humanities\cite{wheeles2013juxta}.
While the strategies found in these works are effective for general use, we are particularly interested in a single dataset. This allows us to specialize our visual encoding to the particular shape of this dataset, as well as to ignore potential scalability issues.
Our design draws on the design lessons highlighted by Correll \etal \cite{correll2012shakespeare}, in that we make text readily available and feature the ability to highlight and investigate outliers.

\section{Case study: Goethe and De Candolle}\label{sec:case}

Our case study concerns the history of scientific ideas in \nth{19}-century Europe.
By the end of the \nth{18}-century, certain stereotypes were formed and prevailed among major European nations to characterize---and to criticize---each other's intellectual atmosphere. For example, France, the center of scientific knowledge during the early modern period, distanced itself from its rival nation, Germany, by criticizing the latter's overly philosophical spirit as inappropriate for science. The impact of the national images of science is strikingly illustrated in the comparison of the reception of Goethe and De Candolle in the French scientific community. Both naturalists described a similar morphological concept, but Goethe was brushed aside as a German philosopher-poet who stumbled upon a scientific concept by chance, while De Candolle, a Swiss naturalist with French lineage, was taken much more seriously.

Traditional close reading of their texts yields evidence both supporting and refuting \nth{19}-century France's image of Goethe's natural science (and of German science more generally).  Yet, the same is true for De Candolle's texts.
Thus, it is not enough to base one's understanding of the given text, and of the author, on a specific part or parts of the text alone. Doing so would easily lead us to the same overly generalized conclusions that historical actors made in the early \nth{19} century.
Thus, we complement close reading with a synthetic reading method that allows us to see the character of the text and the author as a whole. In this analysis we sought to understand the extent to which \nth{19}-century national images of science were supported by the actual scientific texts written by the authors of different nationalities.
%
%
Do the texts relevant to this project belong to different styles of scientific thought? What do their differences really consist of---do the overall characters of the texts match up to the typical images of French or German styles, as was argued by the \nth{19}-century scientists? Or were they independent of the purported national styles of thought?


\subsection{Data and Classification Process}\label{sec:data}

We sought to answer these questions through a near-by reading of several representative scientific texts by Goethe and De Candolle. For Goethe, \textit{Metamorphosis of Plants} (originally published in 1790, translated into French in 1831, denoted in our application and in this paper as \textbf{Goethe});  for De Candolle, \textit{Essay on the Medical Properties of Plants} (1804, \textbf{DC1}), \emph{Elementary Essay on Botanical Geography} (1820, \textbf{DC2}), and a chapter from \textit{Organography of Plants} (1827, \textbf{DC3}).
The French editions of these texts were translated into English and split into sentences.
\textbf{Goethe} contained 382 sentences, \textbf{DC1} had 374, \textbf{DC2} had 800, and \textbf{DC3} had 79.
In \nth{19}-century French, conjoining multiple sentences with semicolons was common. We account for this practice in our classification, c.f. \tabref{tab:categories}.

These texts were then analyzed through a manual categorization constructed through an iterative coding scheme executed by one of the authors who is expert in the field.
These categories were not determined prior to, but during the initial close reading of the texts to maximize their relevance to this study.
Several authoritative \nth{19}-century sources were consulted to establish some of the key categories (i.e. \emph{Romantic}\Rom{} and \emph{Classical}\Clas{}).

The tagging process was repeated eight times over all texts by one of the authors. After the fifth round of tagging, two historians from our university---but outside of the project-team---were invited to apply our scheme to sample paragraphs to verify the replicability of the classification. They were provided with a detailed description of each category before conducting their own classifications.
Their results exhibited a high degree of overlaps with our classification.
After comparing their classifications with our own, we made a necessary revision and conducted two more rounds of tagging process, after which the results showed no difference between the previous and the final rounds.
In the final categorization, these sentences had an average of 3 tags, with a maximum of 5 and minimum of 1.
The resulting category scheme is found in \tabref{tab:categories}.
%
The manual categorization of sentences may seem to diminish the objectivity of the project. However, we believe this freedom of manually selecting and defining classification categories is advantageous. It constitutes a particular researcher's expert-perspective, fitting for history and other fields in humanities, and provides an ideal basis on which to exhibit the intermixing of close and distant reading.

\subsection{Application}\label{sec:application}

\begin{figure}[t] 
  \centering
  \includegraphics[width=\linewidth]{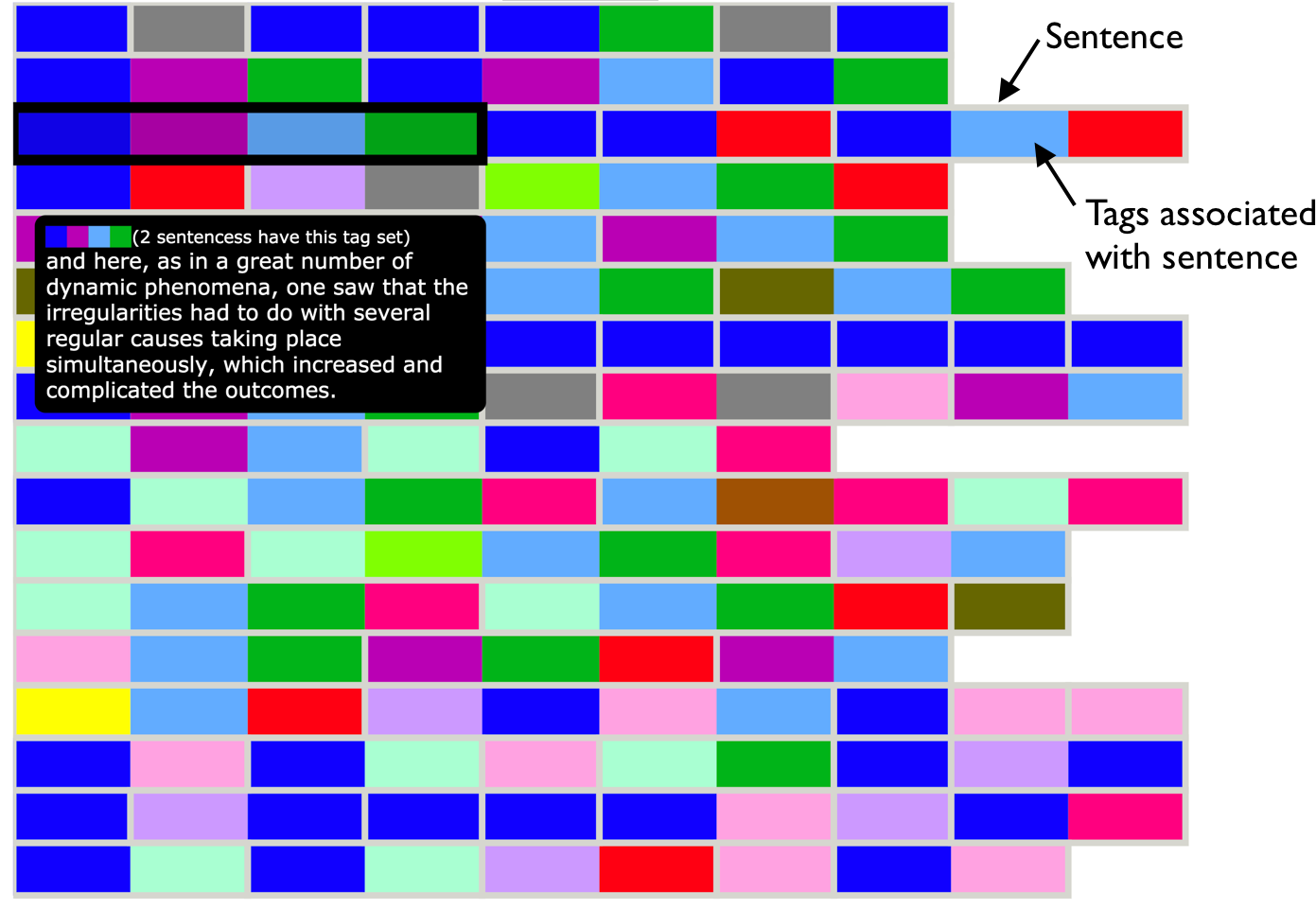}
  \caption{Our waffle view showing \textbf{DC3}. Each tag is denoted as an equally sized and appropriately colored rectangle. Sentences are shown in text order as unfilled boxes containing their tags.}
  \label{fig:waffle}
\end{figure}

To answer our research questions we created a web-based visual analysis application that provides a rich view of the multi-tag classification structure present in our derived dataset.
We show a view of our application in \figref{fig:application}.
Texts can be filtered to remove overly dominant tags or outliers, or to enable comparison of a single text with a subset of that text.
Both columns---each of which corresponds to a single text---can display one of three visualizations: a \textbf{graph} view (\figref{fig:application} right), which enables set-like comparisons of tags, a \textbf{matrix} view (\figref{fig:application} left), which enables the identification of correlation between categories, and a \textbf{waffle} view (\figref{fig:waffle}), which displays tag chronology across the text.
Each view also shows a simple visual summary of the current dataset by a pair of bar charts.
This system serves both as a medium and presentation of analysis, as it is a piece of a larger page describing our project on \nth{19}-century scientific texts.
We now describe each of these views in detail.

\begin{figure}[t] 
  \centering
  \includegraphics[width=\linewidth]{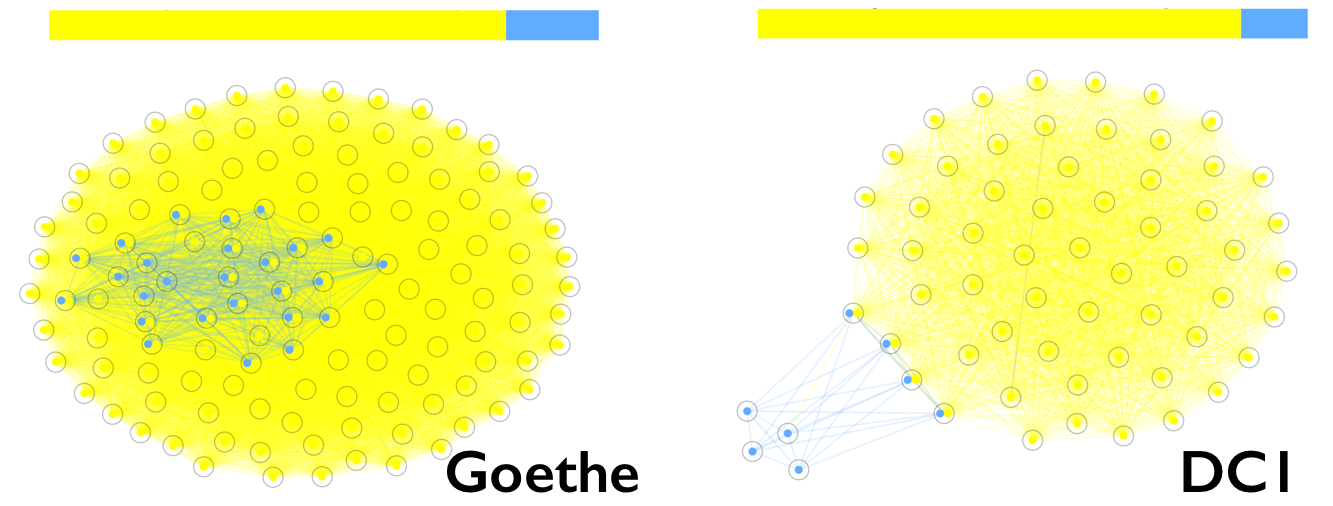}
  \caption{
    Simple summary graphics, such as stacked bar charts (top), indicate that the ratio of \emph{Empirical}\Emp{} to  \emph{Romantic}\Rom{} writing in  \textbf{Goethe} and \textbf{DC1} is roughly the same.
    However, the graphs reveal more nuance in each author's relationship with those modes of writing.
    For \textbf{Goethe}, \emph{Romantic}\Rom{} is completely encompassed by \emph{Empirical}\Emp{}, while in \textbf{DC1}, there is a clear separation between the classes.
  }
  \label{fig:visual-1}
\end{figure}

\begin{figure}[t] 
  \centering
  \includegraphics[width=\linewidth]{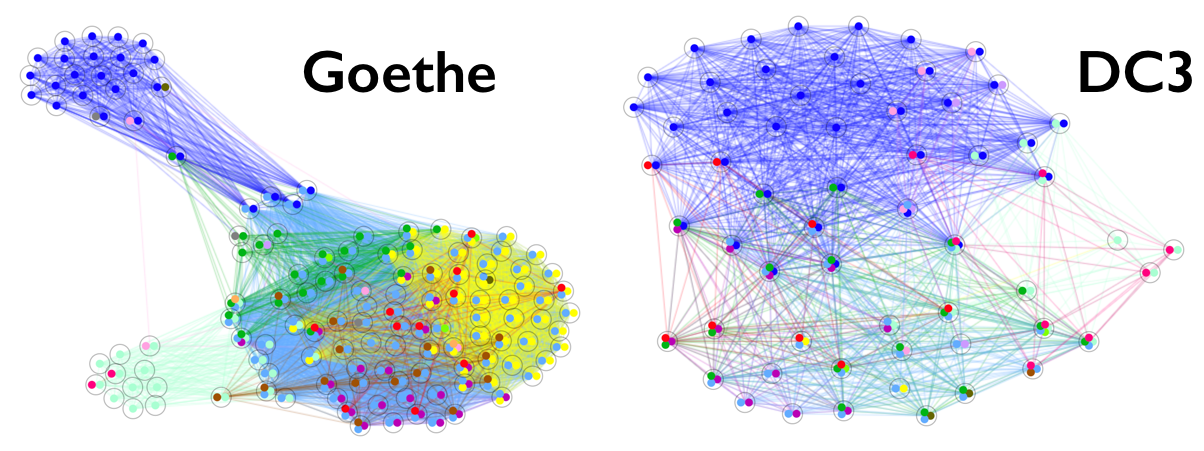}
  \caption{
    The role of \emph{Romantic}\Rom{} and \emph{Classical}\Clas{} writing
    are critical to our analysis, as the \nth{19}-century French scientists tended to accept an image of the German science as \emph{Philosophical}\Phil{} and \emph{Romantic}\Rom{},
    in contrast to their self-image of being \emph{Empirical}\Emp{} and \emph{Classical}\Clas{}.
  }
  \label{fig:visual-3}
\end{figure}

\parahead{Graph}
Our graph view (\figref{fig:application} right, \figref{fig:visual-1}, \figref{fig:visual-3}) visualizes the relationship between the multi-tag classifications through a dimensionality reduction scheme.
We observed that the maximum number of categories for a single sentence was low (five for \textbf{DC3}). Following this observation we created a glyph to denote each sentence, treated graphically as a node. We then link each of these nodes together, such that all nodes that share a tag are connected. This exposes a graphical representation of each category as the sum of the lines denoting it and therein highlighting regions of union and disjunction.
Users can hover over each node to see the corresponding sentence and how many other sentences share that tag combination.
%
%
We construct this layout by treating the categorization of each sentence as a 17-dimensional vector, which we embed into two-dimensions via t-SNE \cite{maaten2008visualizing}. The embedding yields a great deal of overlapping points, which we reduce via force-direction.
%
This creates a visual approximation to the set comparisons of Euler diagrams, without encountering the difficulties traditionally associated with set drawing.
The resulting image gives access to the relationships within the categories, which, as we discuss in \secref{sec:insights}, allows for analysis of the interplay between modes of writing across the texts.
This visual form matches the non-spatial nature of our data, and are a  familiar form of visualization in distant reading \cite{moretti2005graphs, stasko2008jigsaw,campbell2018close,janicke2017visual, finegold2016six}.

\parahead{Matrix}
Our matrix view (\figref{fig:application} left) is the most traditional of our visualizations. It shows the co-occurrence of each category across the chosen text as a heatmap.  This naturally augments our graph view as matrix views are often used as a dual representation to the node-link diagram. Digital humanities projects have used matrix displays to show various relationships \cite{ruecker2005interface, janicke2017visual}.

\parahead{Waffle}
Finally, our waffle view (\figref{fig:waffle}) offers a variation on waffle plots adapted to describe hierarchical ordered groups---in our case, ordered multi-tagged sentences. Each sentence is represented as a variably sized rectangle, made up of equal sized rectangles denoting the tags associated with that sentence. Sentences are arranged from left to right. As in the graph view, users can hover over to view the corresponding sentences, as well as the number of other sentences that share that tag combination. This view gives visual primacy to the categories over the sentences, enabling the viewer to see trends across the text and the relative distribution of tags for that text.

\subsection{Insights Gained from this Approach}\label{sec:insights}



To illustrate how this method benefited our historical project---and therein offer validation of our system and approach---we now describe some of the key insights gained through its application.



We sought to test the commonly accepted view that Goethe's (and German) natural science was far less empirical than that of the French. This suggested that it was essential to understand \emph{Empirical's}\Emp{} relationships with other categories.

A close reading only reveals what the bar charts indicate in \figref{fig:visual-1}---that both authors use similar percentages of their texts to discuss their empirical findings. This knowledge alone would successfully contradict the hypothesis we were testing. Through our near-by reading method, however, we gained a deeper, unexpected knowledge.

In \figref{fig:visual-1} we see that there is a strong correlation between \emph{Romantic}\Rom{} and \emph{Empirical}\Emp{} categories in \textbf{Goethe}, compared to their separation in \textbf{DC1}. This suggests that the two authors treated and expressed their empirical findings in different manners. Goethe's experience of nature was directly translated through Romantic language, while De Candolle treated Romantic language as a conceptual system that a scientist would choose to utilize at specific places.

Thus, our method led us to a conclusion that the real difference between the authors had nothing to do with the quantity of empirical data, but was really about \textit{how} each author expressed their empirical contents throughout the texts. The difference in the manner of expressing scientific information is a subtle yet important point.

Next, in \figref{fig:visual-3}, we see that 
much of \emph{Romantic}\Rom{} and \emph{Classical}\Clas{} elements of the text are absorbed by \emph{Methodological}\Method{} and \emph{Historical}\Hist{} purposes in De Candolle.
In contrast, for \textbf{Goethe}, Romantic and Classical elements of language have greater distinction. This writing is visually and textually distinct from the historical and methodological writing. In \textbf{Goethe}, Romantic elements tend to be associated with the categories that directly involve the observer (ex. \emph{Empirical}\Emp{}). In \textbf{DC3}, they are part of a methodological or historical system for understanding nature.
Understanding data (textual or otherwise) from a distance can offer opportunities for serendipitous discovery of insights \cite{alexander2014serendip,thudt2012bohemian}. Beyond simply confirming hypotheses garnered from close reading, applying visual analytics methods as a part of a near-by reading can enable just this sort of fortuitous discovery. Prior to our study we believed that \emph{Historical}\Hist{} and \emph{Methodological}\Method{} classes would be of little interest. However, our visuals highlighted their interaction with some of our main tags of interest--- \emph{Romantic}\Rom{} and \emph{Classical}\Clas{}---in a markedly different manner in the two texts being compared.


Under either a solely close or distant reading, we would not have found these finer points.
These insights support the overall argument that the images of national scientific styles played a significant role in how \nth{19}-century France received foreign works, and successfully identify the real points of differences that had very little to do with the assumed national differences.
These analyses illustrate how near-by readings can guide the researcher into given texts and minds of their authors.
While these visualizations might be understood by a reader who is not familiar with the texts--a distant reading--these images are artifacts of the near-by reading described here, and cannot be meaningfully disentangled.

\section{Discussion}



%

In this paper we introduced the notion of near-by reading---an idea founded on the premise that close and distant reading form a spectrum---and provided a case study in which such a perspective would be valuable. This study demonstrated the utility of a mixed digital reading. Its bespoke visualizations, while not necessarily novel, enhance an expert's ability to understand and compare a set of texts, which allow for identifying contradictions, both in terms of outliers \cite{hope2010hundredth, correll2012shakespeare}, and, as in our study, within broader historical contexts.
By reconciling close and distant reading, individual interpretation and externalized computational analysis, our near-by reading allowed us to contribute to the history of \nth{19}-century biology some new insights that would have remained invisible under other techniques.




%

Yet all is not necessarily rosy for near-by reading. Eve rightly points out that these analyses are subject to HARKing (hypothesizing after results are known) \cite{eve2019close}. Given the malleable notion of what constitutes data in historical investigations, investigating scholars can easily run into \emph{visualization mirages}, or significant insights that disappear upon closer inspection\cite{mcnuttt2020mirage}.
Da provides a litany of problems that can arise when statistical techniques are improperly applied or interpreted \cite{da2019computational}.
Computational analyses and visualizations can possess an rhetorically domineering quality\cite{drucker2011humanities}, inspiring trust in uncritical readings, where they should be closely questioned and carefully investigated \cite{mcnutt2020divining}.
The design of systems that augment expertise\cite{heer2019agency} without giving way to ephemeral or untrustworthy insights is a critical problem in the design of systems that politely \cite{whitworth2005polite} navigate the space of exchanging agency for computational power. This problem is even more difficult in the space of the digital humanities where notions of ground truth---and therein accuracy of analysis---can be epistemologically hazy \cite{correll2012shakespeare}.
Some of these concerns can be addressed if, returning to our definition of the near-by reading, researchers are familiar with the texts on hand to some extent.
By placing the distanced component of our reading on a close foundation, we have a higher probability to stray away from such treacherous analysis pitfalls.

Near-by reading exists among a universe of other possible reading combinations (e.g. conducting distant reading before getting acquainted with each text, mechanically alternating back and forth between near and far) that serve other purposes and effects.
Our particular instantiation is suitable for scholars who are concerned with specific \textit{comparative} points across modestly-sized collections of texts and who want to have an equal hand on the analysis as the computational prostheses \cite{hope2010hundredth}.



Keeping these cautions and aims in mind, we believe that near-by reading can offer real utility for both historical projects such as our own and for other humanities projects more generally.
Such mixed digital and analog approaches to scholarship can birth novel hypotheses that can be verified with further close reading.
In turn these readings can be useful in sparking previously overlooked questions in humanities research, thereby helping experts identify the types or areas of sources that necessitate greater study.
A humanities discipline like history necessarily deals with the figures and events that are no longer active, and to prevent the historical discussions from ossifying, it is imperative to bring in new perspectives and questions.
These approaches can provide support for existing hypotheses by offering more concrete evidence than a researcher's purely personal interpretation.
We believe that there is an exciting landscape of opportunities to be explored using these techniques.

\acknowledgments{
  The authors wish to thank Will Brackenbury, Ashley Clark, Michael Correll, John A. Goldsmith, Biying Ling, Robert J. Richards, and Michael Rossi. This project received the support of the Visualization for Understanding and Exploration project at the Neubauer Collegium and the Research Computing Cluster of the University of Chicago.}

\bibliographystyle{abbrv-doi}

\bibliography{nearby-bib}

\begin{thebibliography}{10}

\bibitem{alexander2014serendip}
E.~Alexander, J.~Kohlmann, R.~Valenza, M.~Witmore, and M.~Gleicher.
\newblock Serendip: Topic model-driven visual exploration of text corpora.
\newblock In {\em 2014 IEEE Conference on Visual Analytics Science and
  Technology (VAST)}, pp. 173--182. IEEE, 2014.

\bibitem{campbell2018close}
S.~Campbell, Z.-Y. Yu, S.~Connell, and C.~Dunne.
\newblock {Close and Distant Reading via Named Entity Network Visualization: A
  Case Study of Women Writers Online}.
\newblock In {\em Proceedings of the 3rd Workshop on Visualization for the
  Digital Humanities. VIS4DH}, 2018.

\bibitem{correll2012shakespeare}
M.~Correll and M.~Gleicher.
\newblock {What Shakespeare Taught Us About Text Visualization}.
\newblock In {\em IEEE Visualization Workshop Proceedings: The 2nd Workshop on
  Interactive Visual Text Analytics: Task-Driven Analysis of Social Media
  Content}, 2012.

\bibitem{correll2011exploring}
M.~Correll, M.~Witmore, and M.~Gleicher.
\newblock {Exploring Collections of Tagged Text for Literary Scholarship}.
\newblock In {\em Computer Graphics Forum}, vol.~30, pp. 731--740. Wiley Online
  Library, 2011. doi: {{%
10\hspace{.1pt}\discretionary{.}{%
}{.}\hspace{.4pt}1111\discretionary{/}{%
}{/}j\hspace{.1pt}\discretionary{.}{%
}{.}\hspace{.4pt}1467\discretionary{%
}{-}{-}8659\hspace{.1pt}\discretionary{.}{%
}{.}\hspace{.4pt}2011\hspace{.1pt}\discretionary{.}{%
}{.}\hspace{.4pt}01922\hspace{.1pt}\discretionary{.}{%
}{.}\hspace{.4pt}x}}


\bibitem{da2019computational}
N.~Z. Da.
\newblock {The Computational Case against Computational Literary Studies}.
\newblock {\em Critical inquiry}, 45(3):601--639, 2019. doi: {{%
10\hspace{.1pt}\discretionary{.}{%
}{.}\hspace{.4pt}1086\discretionary{/}{%
}{/}702594}}


\bibitem{drucker2011humanities}
J.~Drucker.
\newblock {Humanities Approaches to Graphical Display}.
\newblock {\em Digital Humanities Quarterly}, 5(1):1--21, 2011.

\bibitem{edelstein2018england}
D.~Edelstein and B.~Kassabova.
\newblock {How England Fell off the Map of Voltaire's Enlightenment}.
\newblock {\em Modern Intellectual History}, pp. 1--25, 2018.

\bibitem{el2016visual}
M.~El-Assady, V.~Gold, M.~John, T.~Ertl, and D.~A. Keim.
\newblock {Visual Text Analytics in Context of Digital Humanities}.
\newblock In {\em Proceedings of the 1st Workshop on Visualization for the
  Digital Humanities. VIS4DH}, 2016.

\bibitem{eve2019close}
M.~P. Eve.
\newblock {\em Close Reading with Computers: Textual Scholarship, Computational
  Formalism, and David Mitchell's Cloud Atlas}.
\newblock 2019.

\bibitem{froehlich2019Moby}
H.~Froehlich.
\newblock {Moby Dick is About Whales, or Why Should We Count Words?}
\newblock https://hfroehli.ch/blog-2/.

\bibitem{gazzoni2017mapping}
A.~Gazzoni.
\newblock {Mapping Dante: A Digital Platform for the Study of Places in the
  Commedia}.
\newblock {\em Humanist Studies \& the Digital Age}, 5(1):82--95, 2017.

\bibitem{heer2019agency}
J.~Heer.
\newblock {Agency Plus Automation: Designing Artificial Intelligence into
  Interactive Systems}.
\newblock {\em Proceedings of the National Academy of Sciences},
  116(6):1844--1850, 2019. doi: {{%
10\hspace{.1pt}\discretionary{.}{%
}{.}\hspace{.4pt}1073\discretionary{/}{%
}{/}pnas\hspace{.1pt}\discretionary{.}{%
}{.}\hspace{.4pt}1807184115}}


\bibitem{hope2010hundredth}
J.~Hope and M.~Witmore.
\newblock {The Hundredth Psalm to the Tune of ``Green Sleeves": Digital
  Approaches to Shakespeare's Language of Genre}.
\newblock {\em Shakespeare Quarterly}, 61(3):357--390, 2010.

\bibitem{janicke2017visual}
S.~J{\"{a}}nicke, G.~Franzini, M.~F. Cheema, and G.~Scheuermann.
\newblock {Visual Text Analysis in Digital Humanities}.
\newblock {\em Computer Graphics Forum}, 36(6):226--250, 2017. doi: {{%
10\hspace{.1pt}\discretionary{.}{%
}{.}\hspace{.4pt}1111\discretionary{/}{%
}{/}cgf\hspace{.1pt}\discretionary{.}{%
}{.}\hspace{.4pt}12873}}


\bibitem{koch2014varifocalreader}
S.~Koch, M.~John, M.~W{\"{o}}rner, A.~M{\"{u}}ller, and T.~Ertl.
\newblock {VarifocalReader - In-Depth Visual Analysis of Large Text Documents}.
\newblock {\em {IEEE} Transactions on Visualization and Computer Graphics},
  20(12):1723--1732, 2014. doi: {{%
10\hspace{.1pt}\discretionary{.}{%
}{.}\hspace{.4pt}1109\discretionary{/}{%
}{/}TVCG\hspace{.1pt}\discretionary{.}{%
}{.}\hspace{.4pt}2014\hspace{.1pt}\discretionary{.}{%
}{.}\hspace{.4pt}2346677}}


\bibitem{lex2014upset}
A.~Lex, N.~Gehlenborg, H.~Strobelt, R.~Vuillemot, and H.~Pfister.
\newblock {UpSet: Visualization of Intersecting Sets}.
\newblock {\em {IEEE} Transactions on Visualization and Computer Graphics},
  20(12):1983--1992, 2014. doi: {{%
10\hspace{.1pt}\discretionary{.}{%
}{.}\hspace{.4pt}1109\discretionary{/}{%
}{/}TVCG\hspace{.1pt}\discretionary{.}{%
}{.}\hspace{.4pt}2014\hspace{.1pt}\discretionary{.}{%
}{.}\hspace{.4pt}2346248}}


\bibitem{maaten2008visualizing}
L.~v.~d. Maaten and G.~Hinton.
\newblock {Visualizing Data Using t-SNE}.
\newblock {\em Journal of Machine Learning Research}, 9(Nov):2579--2605, 2008.

\bibitem{mccurdy2015poemage}
N.~McCurdy, J.~Lein, K.~Coles, and M.~Meyer.
\newblock {Poemage: Visualizing the Sonic Topology of a Poem}.
\newblock {\em {IEEE} Transactions on Visualization and Computer Graphics},
  22(1):439--448, 2015. doi: {{%
10\hspace{.1pt}\discretionary{.}{%
}{.}\hspace{.4pt}1109\discretionary{/}{%
}{/}TVCG\hspace{.1pt}\discretionary{.}{%
}{.}\hspace{.4pt}2015\hspace{.1pt}\discretionary{.}{%
}{.}\hspace{.4pt}2467811}}


\bibitem{mcnutt2020divining}
A.~McNutt, A.~Crisan, and M.~Correll.
\newblock {Divining Insights: Visual Analytics Through Cartomancy}.
\newblock In {\em Extended Abstracts of the 2020 CHI Conference on Human
  Factors in Computing Systems}, pp. 1--16, 2020.

\bibitem{mcnuttt2020mirage}
A.~McNutt, G.~Kindlmann, and M.~Correll.
\newblock {Surfacing Visualization Mirages}.
\newblock In {\em Proceedings of the 2020 CHI Conference on Human Factors in
  Computing Systems}, CHI ’20, p. 1–16. ACM, 2020. doi: {{%
10\hspace{.1pt}\discretionary{.}{%
}{.}\hspace{.4pt}1145\discretionary{/}{%
}{/}3313831\hspace{.1pt}\discretionary{.}{%
}{.}\hspace{.4pt}3376420}}


\bibitem{moretti2005graphs}
F.~Moretti.
\newblock {\em Graphs, Maps, Trees: Abstract Models for a Literary History}.
\newblock Verso, 2005.

\bibitem{riche2010untangling}
N.~H. Riche and T.~Dwyer.
\newblock {Untangling Euler Diagrams}.
\newblock {\em {IEEE} Transactions on Visualization and Computer Graphics},
  16(6):1090--1099, 2010. doi: {{%
10\hspace{.1pt}\discretionary{.}{%
}{.}\hspace{.4pt}1109\discretionary{/}{%
}{/}TVCG\hspace{.1pt}\discretionary{.}{%
}{.}\hspace{.4pt}2010\hspace{.1pt}\discretionary{.}{%
}{.}\hspace{.4pt}210}}


\bibitem{ruecker2005interface}
S.~Ruecker, S.~Ramsay, M.~Radzikowska, and A.~Galey.
\newblock {Interface Design}.
\newblock {\em Proceedings of the Digital Humanities}, 7(8):11, 2005.

\bibitem{stasko2008jigsaw}
J.~Stasko, C.~G{\"o}rg, and Z.~Liu.
\newblock {Jigsaw: Supporting Investigative Analysis through Interactive
  Visualization}.
\newblock {\em Information Visualization}, 7(2):118--132, 2008. doi: {{%
10\hspace{.1pt}\discretionary{.}{%
}{.}\hspace{.4pt}1057\discretionary{/}{%
}{/}palgrave\hspace{.1pt}\discretionary{.}{%
}{.}\hspace{.4pt}ivs\hspace{.1pt}\discretionary{.}{%
}{.}\hspace{.4pt}9500180}}


\bibitem{thudt2012bohemian}
A.~Thudt, U.~Hinrichs, and S.~Carpendale.
\newblock The bohemian bookshelf: supporting serendipitous book discoveries
  through information visualization.
\newblock In {\em Proceedings of the SIGCHI Conference on Human Factors in
  Computing Systems}, pp. 1461--1470, 2012.

\bibitem{finegold2016six}
C.~N. Warren, D.~Shore, J.~Otis, L.~Wang, M.~Finegold, and C.~R. Shalizi.
\newblock {Six Degrees of Francis Bacon: A Statistical Method for
  Reconstructing Large Historical Social Networks}.
\newblock {\em Digital Humanities Quarterly}, 10(3), 2016.

\bibitem{wheeles2013juxta}
D.~Wheeles and K.~Jensen.
\newblock {Juxta Commons}.
\newblock In {\em 8th Annual International Conference of the Alliance of
  Digital Humanities Organizations, {DH} 2013, Lincoln, NE, USA, July 16-19,
  2013, Conference Abstracts}, p. 545. Alliance of Digital Humanities
  Organizations {(ADHO)}, 2013.

\bibitem{whitworth2005polite}
B.~Whitworth.
\newblock {Polite Computing}.
\newblock {\em Behaviour \& Information Technology}, 24(5):353--363, 2005. doi:
  {{%
10\hspace{.1pt}\discretionary{.}{%
}{.}\hspace{.4pt}1080\discretionary{/}{%
}{/}01449290512331333700}}


\end{thebibliography}
\end{document}